\documentclass[twocolumn,amssymb,aps,prd]{revtex4}
\usepackage{amsmath,amssymb}

\newcommand{\fr}[1]{\frac{1}{#1}}

\newcommand{\ord}[1]{{\mathcal O}\left(#1\right)}

\newcommand{\eps}{\epsilon}
\newcommand{\de}{\delta}
\newcommand{\be}{\beta}
\newcommand{\la}{\lambda}
\newcommand{\ga}{\gamma}
\newcommand{\teps}{\Tilde \eps}
\newcommand{\ro}{\rho}

\usepackage{graphicx}

\begin{document}
\title{Extreme charged black holes in braneworld with cosmological constant}

\author{Ryotaku Suzuki}
\email{ryotaku@tap.scphys.kyoto-u.ac.jp},
\affiliation{Department of Physics, Kyoto University, Kyoto 606-8502, Japan}
\author{Tetsuya Shiromizu}
\email{shiromizu@tap.scphys.kyoto-u.ac.jp},
\affiliation{Department of Physics, Kyoto University, Kyoto 606-8502, Japan}
\author{Norihiro Tanahashi}
\email{tanahasi@yukawa.kyoto-u.ac.jp},
\affiliation{Yukawa Institute for Theoretical Physics, Kyoto University, Kyoto 606-8502, Japan}


\begin{abstract}
Application of the adS/CFT correspondence to the RS models may predict
that there is no static solution for black holes with a radius larger than the bulk curvature scale. 
When the black hole has an extremal horizon, however, the correspondence 
suggests that the black hole can stay static.
We focus on the effects of cosmological constant on the brane on such extremal brane-localized black holes.
We observe that the positive cosmological constant restrict the black hole size on the brane as in ordinary four-dimensional general relativity.
The maximum black hole size differs from that in four-dimensional general relativity case 
due to the non-linear term in the effective Einstein equation.
In the negative cosmological constant case, we obtain an implication on the Newton constant in the Karch-Randall model.
\end{abstract}

\maketitle

\section{Introduction}
The braneworld model is a phenomenological model which describes our four-dimensional universe
in higher-dimensional theory.
In this model, we are living on a four-dimensional membrane, 
and only gravity propagates to the extra dimension.
Among several models for braneworlds, Randall-Sundrum (RS) type models are 
 interesting because they provide 
us many phenomenological predictions~\cite{RS99a,RS99b,MK10}.
In these models, extra dimension
is warped due to the self-gravity of the branes.
Because of this warping,
it is found in some RS type models 
that 
gravity can be confined near the brane and becomes four-dimensional
even when the extra dimension is non-compact~\cite{GT00,KR01}. 

Although many studies on the RS model have been done, there are still some open 
issues. 
One of them is that 
static solutions of black holes localized on the brane are missing.
Though numerical solutions of such brane-localized black holes are constructed when the 
black hole size is smaller than the bulk curvature scale~\cite{KTN03,Yos09}, 
no solutions are found 
when the size is large.
For this issue, the following conjecture 
has been proposed based on the adS/CFT 
correspondence~\cite{Tan02,EFK02,EGK02}~(see also Ref.~\cite{Tana} for related issues). 
According to the correspondence, a five-dimensional 
classical brane-localized black hole is dual to 
a four-dimensional black hole that emits the Hawking radiation. 
Since the latter one cannot be static due to the Hawking radiation emission, 
it is suggested by the duality that there is no static brane-localized black 
hole which is larger than the bulk curvature radius.

Here, one might realize that the adS/CFT correspondence also tells 
that static solutions may present when the black hole horizon is extreme~\cite{KR09}
since
the horizon temperature is zero and the Hawking radiation will not be emitted.
Indeed, the authors of Ref.~\cite{KR09} constructed
the near-horizon geometry of such extreme charged static black hole 
localized on the asymptotically flat brane and studied its properties.
In this paper, we shall consider the near-horizon geometry of extreme charged 
black hole localized on the brane with non-vanishing cosmological constant
to study the properties of the brane-localized black holes in more generalized settings.
We also intend to reveal the non-trivial property of the gravity 
in the braneworld model with negative cosmological constant, the Karch-Randall model.

The rest of this paper is organized as follows. 
In Sec.~\ref{sec:model}, we describe the model we study. 
We sketch the metric form for the near-horizon geometry in Sec.~\ref{sec:setup}
and we present numerical solutions in Sec.~\ref{sec:solutions}.
In Sec.~\ref{sec:largebh}, we give analytic arguments for relatively large black holes. 
Finally, we give summary and discussion in Sec.~\ref{sec:summery}.

\section{Models} \label{sec:model}

The model we consider in this paper is the RS braneworld model, which consists of 
five-dimensional asymptotically anti-de Sitter (adS) bulk spacetime and 
a four-dimensional brane with positive tension in it.
The action of this model is given by
\begin{eqnarray}
S & = & \fr{2\kappa^2_5}\int_M d^5x\sqrt{-g}  \left( {}^{(5)} R + \frac{12}{l^2}\right) 
\nonumber \\
& & +\fr{\kappa^2_5}\int_{\partial M}d^4x \sqrt{-h}  K \nonumber \\
& & +\int_{\rm brane} d^4x \sqrt{-h} \left(-\sigma -\fr{2\kappa^2_4}F_{\mu\nu}F^{\mu\nu}\right), 
\end{eqnarray}
where $M$ is the bulk spacetime and $\partial M$ is its outer boundary. 
$h_{\mu\nu}$ is the induced metric on the brane. $\kappa_5^2=8\pi G_5$ and 
$\kappa_4^2=8 \pi G_4$ are the five and four-dimensional 
gravitational coupling, 
respectively. $l$ is the bulk curvature radius. 
$\sigma$ and $F_{\mu\nu}$ are the brane tension and 
the field strength of the Maxwell field on the brane. 
$K$ is the trace of the extrinsic curvature $K_{\mu\nu}$ of $\partial M$. We 
impose the $Z_2$-symmetry about the brane. 

From the above action, we obtain the five-dimensional Einstein equation in the 
bulk as 
\begin{eqnarray}
R_{MN}-\fr{2}Rg_{MN}=\frac{4}{l^2}g_{MN}. \label{eq:5deinstein}
\end{eqnarray}
Under the $Z_2$-symmetry, the Israel's junction condition on the 
brane is given by~\cite{Isr66}
\begin{eqnarray}
K_{\mu\nu}-K h_{\mu\nu} = \frac{1}{2}\kappa_5^2 T_{\mu\nu}, \label{eq:israeljc}
\end{eqnarray}
where $T_{\mu\nu}$ is the energy-momentum tensor on the brane, which is given as 
\begin{eqnarray}
T_{\mu\nu} = - \sigma h_{\mu\nu} + \frac{2}{\kappa_4^2}\left(F_{\mu \alpha}{F_\nu}^\alpha
-\fr{4}F^2 h_{\mu\nu}\right).
\end{eqnarray}
The Maxwell equation and the Bianchi equation are
\begin{eqnarray}
d * F =0,
\quad
dF=0,
\end{eqnarray}
where * is the Hodge dual in four dimensions.

\section{Near-horizon geometry, bulk equations and boundary conditions} \label{sec:setup}

\subsection{Near-horizon geometry}

We consider a static brane-localized black hole whose horizon is 
made to be extreme by the Maxwell field on the brane.
A static black hole has constant surface gravity on its horizon.
Then, when the horizon is extremal 
on the intersection with the brane,
the whole part of the horizon in the bulk will also be extremal.
For such an extremal horizon, we can take the near-horizon limit and analyze its properties.
It is proved that the near-horizon geometry of a static extreme black hole can be 
written in a warped product of a two-dimensional Lorentzian space and a compact 
manifold as~\cite{KLR07}
\begin{eqnarray}
ds^2=A(x)^2d\Sigma^2+g_{ab}dx^a dx^b, 
\end{eqnarray}
where $d\Sigma^2$ is a two-dimensional Lorentzian metric 
$M_2$ of constant curvature $2k$. When the metric describes the 
black hole spacetime, $k$ should be negative and then $M_2$ is two-dimensional 
AdS spacetime (${\rm adS}_2$). 
We also assume that $g_{ab}dx^a dx^b$ has $SO(3)$ symmetry.
Choosing the coordinates $x^a = (\rho, \theta, \phi)$, the near-horizon geometry becomes
\begin{eqnarray}
ds^2=A(\rho)^2d\Sigma^2 + d\rho^2 + R(\rho)^2d\Omega^2, \label{metric}
\end{eqnarray}
where $d\Omega^2$ is the metric of the two-dimensional unit sphere. 

\subsection{Bulk equations}

For the metric ansatz of Eq.~(\ref{metric}),
the bulk Einstein equations, Eq.~(\ref{eq:5deinstein}), 
becomes
\begin{equation}
	\frac{k}{A^2}-\frac{{A'}^2}{A^2}-\frac{2A'R'}{AR}-\frac{A''}{A}=-\frac{4}{l^2}, 
\label{eom1}
\end{equation}
\begin{equation}
\frac{A''}{A}+\frac{R''}{R}=\frac{2}{l^2} 
\label{eom2}
\end{equation}
and
\begin{equation}
\frac{1}{R^2}-\frac{{R'}^2}{R^2}-\frac{2A'R'}{AR}-\frac{R''}{R}=-\frac{4}{l^2}, 
\label{eom3}
\end{equation}
where prime stands for the derivative with respect to $\rho$. 
From these we obtain 
\begin{eqnarray}
\frac{k}{A^2}+\frac{1}{R^2}=\frac{{A'}^2}{A^2}+\frac{{R'}^2}{R^2}+\frac{4A'R'}{AR}-\frac{6}{l^2},
\label{Hami}
\end{eqnarray}
which is the Hamiltonian constraint. 

We assume the horizon to be compact, which implies that $R(\rho)$ vanishes somewhere.
Then, we set the ``origin'' of $\rho$ as $R(\rho=0)=0$. 
The smoothness of the horizon at the ``origin'' requires $R'(0)=1$ and $A'(0)=0$. 
Then, the only free parameter under the boundary condition at $\rho=0$
is $A(0)=A_0$. 
After all, the bulk equations 
have three free parameters \{$A_0,k,l$\}.

Here, note that the equations have two families of scaling invariance: 
$(A,k) \rightarrow (\lambda_1A,\lambda_1^2k)$ and $
(R,l,\rho,k) \rightarrow (\lambda_2R,\lambda_2l,\lambda_2\rho,\lambda_2^{-2}k)$.
Then, we can set $A_0=1$ and $l=1$ without loss of generality,
and 
then,
the only free parameter will be $k$.
After getting a solution $(\tilde{A}(\tilde{\rho}),\tilde{R}(\tilde{\rho}))$, 
we can recover a dimensionful solution as 
$( A_0 \tilde{A}(l^{-1}\rho),l\tilde{R}(l^{-1}\rho))$.

\subsection{Junction condition}

From Eq.~(\ref{eq:israeljc}), the junction condition determines 
the extrinsic curvature $K_{\mu\nu}$ on the brane as 
\begin{eqnarray}
K_{\mu\nu}|_{\rm brane}=\frac{\kappa_5^2\sigma}{6}h_{\mu\nu}
+\frac{\kappa_5^2}{\kappa_4^2}
\left(F_{\mu\alpha}{F_\nu}^\alpha-\frac{1}{4}F^2h_{\mu\nu}\right). 
\label{eq:israeljc2}
\end{eqnarray}
The induced cosmological constant on the brane $\Lambda_4$ is given as~\cite{SMS00}
\begin{eqnarray}
\Lambda_4 \equiv -\frac{3}{l^2}+\frac{\kappa_5^4 \sigma^2}{12}. \label{eq:lambda4}
\end{eqnarray}
From this expression, however, we see that 
$\Lambda_4$ 
is
bounded from below as $\Lambda_4 \geq -3/l$. 
In Ref.~\cite{KR09}, the brane tension is tuned to make the brane 
asymptotically flat.
In our current paper, we will not impose such tuning. 
Then, the brane geometry will be asymptotically
de Sitter, anti-de Sitter or Minkowski spacetimes depending the value of 
$\Lambda_4$.
For convenience,  
we introduce the following dimensionless parameter 
\begin{equation}
\alpha \equiv \frac{\sigma}{\sigma_{RS}},
\end{equation}
where $\sigma_{RS} \equiv 6/\kappa_5^2l$ is the value of the tension when the brane 
geometry is asymptotically Minkowski spacetime. 
$\alpha=1$ corresponds to $\Lambda_4=0$. By the 
definition of $\alpha$ and $\Lambda_4$, 
they are related as 
\begin{eqnarray}
\alpha=\sqrt{1+\frac{l^2\Lambda_4}{3}}=\frac{l\kappa_5^2\sigma}{6}.
\label{eq:alpha}
\end{eqnarray}
Note that $\alpha >1~(\alpha<1)$ for $\Lambda_4 >0~(\Lambda_4 <0)$.

Now, we suppose that the brane is located at $\rho= \rho_0$.  
Then the Israel junction condition~(\ref{eq:israeljc2}) becomes
\begin{equation}
\frac{A(\rho_0)'}{A(\rho_0)}=\alpha-\frac{\kappa_5^2}{\kappa_4^2} \frac{Q^2}{2L_2^4},
\quad
\frac{R(\rho_0)'}{R(\rho_0)}=\alpha+\frac{\kappa_5^2}{\kappa_4^2} \frac{Q^2}{2L_2^4}.
\label{eq:israeljc3}
\end{equation}
Here, we used a notation for the induced metric on the brane such that 
\begin{eqnarray}
ds^2_{\rm brane}=|k|L_1^2d\Sigma^2+L_2^2d\Omega^2, 
\end{eqnarray}
where $L_1$ and $L_2$ are proper radii of $M_2$ and $S^2$ defined by 
\begin{equation}
L_1^2 \equiv {|k|}^{-1}A(\rho_0)^2,
\quad
L_2^2 \equiv R(\rho_0)^2. 
\end{equation}
Moreover, we 
used the solution for the Maxwell field 
\begin{eqnarray}
*F = Qd\Omega, 
\end{eqnarray}
where $Q$ is the total charge on the brane given by
\begin{eqnarray}
Q=\frac{1}{4\pi} \int_{S^2} *F.
\end{eqnarray}

From Eq.~(\ref{Hami}) and the junction condition, we have 
\begin{eqnarray}
\frac{\text{sign}(k)}{L_1^2}+\frac{1}{L_2^2}
&=&
\frac{6}{l}(\alpha^2-1)
-\frac{\kappa_5^4}{\kappa_4^4}\frac{Q^4}{2L_2^8}
\nonumber \\
&=&2\Lambda_4-\frac{\kappa_5^4}{\kappa_4^4}\frac{Q^4}{2L_2^8},
\label{eq:LL}
\end{eqnarray}
where $\text{sign}(k)$ is equal to 1, 0 or $-1$ when $k$ is positive, zero or negative.
From Eq.~(\ref{eq:LL}),
we find some restrictions on the near-horizon geometry.
When $\alpha \leq 1 $, $k$ is always negative. When $\alpha > 1$, on the other hand, 
$k$ can be positive for some large enough values of $\alpha$.

\subsection{Gravitational couplings}
\label{Sec:g-coupling}

Here, we would like to make
a comment on the relation between the four and five-dimensional gravitational couplings. 
From several analyses~\cite{Gen02}, it is sure that the relation for the cases with $\Lambda_4 \geq 0$
($\alpha \geq 1$) is given by 
\begin{eqnarray}
\kappa_4^2=\frac{\kappa_5^4 \sigma}{6}=\frac{\kappa_5^2 \alpha}{l}. 
\end{eqnarray}
On the other hand, we do not have a definite answer for the case of $\Lambda_4 <0$. 
This case is called Karch-Randall model. 
When ${\rm adS}_4$ curvature radius scale is sufficiently larger than 
the bulk curvature scale $l$,
however,
it is expected 
that $\kappa_4^2 \approx \kappa_5^2/l$ holds \cite{KS05}. For the moment, we will use 
the relation $\kappa_4^2=\kappa_5^2/l$ for all ranges of $\Lambda_4$. 
We will ask this issue again in Sec.~\ref{Sec:GC-for-AdS}.

\section{The solutions}\label{sec:solutions}

Let us solve the bulk equations from $\rho=0$ to $\rho=\rho_0$ 
for fixed values of $k$.
In this section, we employ the unit of $l=1$ and also set $A_0=1$.
From the Israel junction condition~(\ref{eq:israeljc3}), 
$Q$ and $\alpha$ are determined as
\begin{eqnarray}
\alpha = \fr{2}\left(\frac{A'}{A}+\frac{R'}{R}\right)|_{\rho=\rho_0} \label{eq:jc-alpha}
\end{eqnarray}
and
\begin{eqnarray}
Q^2 = \frac{\kappa_4^2}{\kappa_5^2}R^4 \left(\frac{R'}{R}-\frac{A'}{A}\right)|_{\rho=\rho_0}. 
\label{eq:jc-Q}
\end{eqnarray}

As shown in Ref.~\cite{KR09}, there are analytic solutions of Eqs.~(\ref{eom1})-(\ref{eom3})
for some special values of $k$. 
One of them is
\begin{eqnarray}
A(\rho)=1, ~~R(\rho)=\fr{\sqrt{2}}\sinh(\sqrt{2}\rho)
\label{eq:exactsol2}
\end{eqnarray}
for $k=-4$.
$k=-1$ yields another exact solution as 
\begin{eqnarray}
A(\rho)=\cosh\rho, ~~R(\rho)=\sinh\rho. \label{exactsol1}
\end{eqnarray}
For these exact solutions, the geometry on the brane is somewhat restricted.
Substituting the above solutions 
into Eq.~(\ref{eq:jc-alpha}), we find
\begin{eqnarray}
\alpha=\fr{\sqrt{2}}\coth(\sqrt{2}\rho_0) 
\stackrel{\rho_0 \to \infty }{=} \fr{\sqrt{2}}, 
\label{former}
\end{eqnarray}
for $k=-4$, and 
\begin{eqnarray}
\alpha=\fr{2}(\tanh\rho_0+\coth\rho_0) >1
\label{latter}
\end{eqnarray}
for $k=-1$. 
From Eq.~(\ref{latter}),
we see that there are no solutions for which $\alpha<1$
when $k=-1$. That is, 
we cannot obtain a brane with $\Lambda_4 \leq 0$ in this case.
On the other hand, when $k=-4$,
we see from Eq.~(\ref{former}) that 
we can realize a brane with $\Lambda_4<0$ if we choose sufficiently large $\rho_0$.

For general $k$, the bulk solution behaves as follows.
\begin{enumerate}
\item $k <-4$ case: $A(\rho)$ monotonically decreases and vanishes at 
a point $\rho_1$. 
At this point, $R(\rho)$ diverges and a curvature singularity appears.
Therefore, the brane position $\rho_0$ must be smaller than $\rho_1$. 
\item $-4 < k <-1$ case: Both $A(\rho)$ and $R(\rho)$ increase exponentially. 
$\alpha$ has a minimum between $1/\sqrt{2}$ and $1$, and tends to $1$ for $\rho_0\to\infty$.
\item $-1 < k$ case: Both $A(\rho)$ and $R(\rho)$ increase exponentially. $\alpha$ decreases 
monotonically and tends to $1$ for $\rho_0\to\infty$.
\end{enumerate}

\begin{figure}[h]
\begin{center}
\includegraphics[width=4cm]{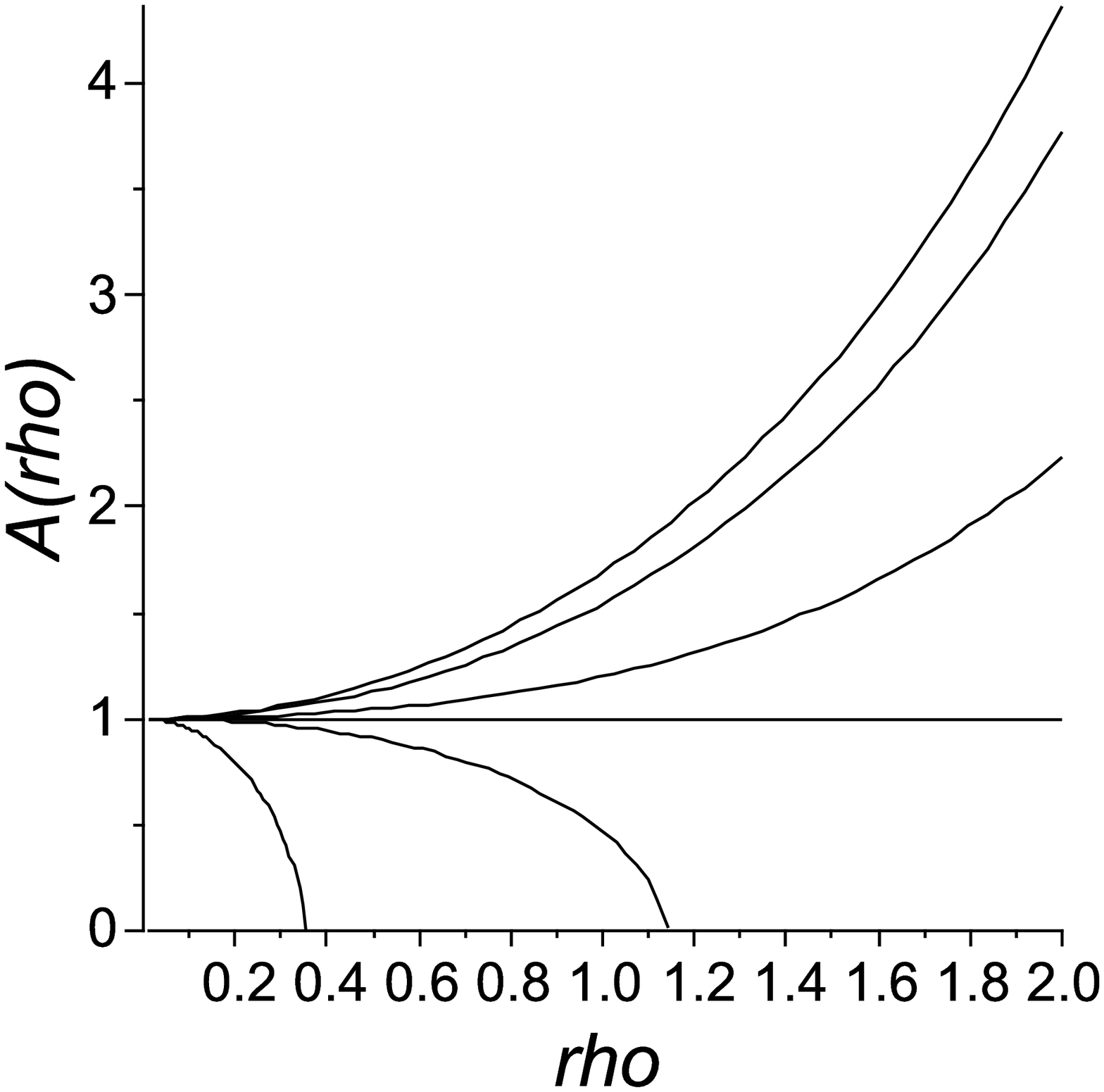}
\includegraphics[width=4cm]{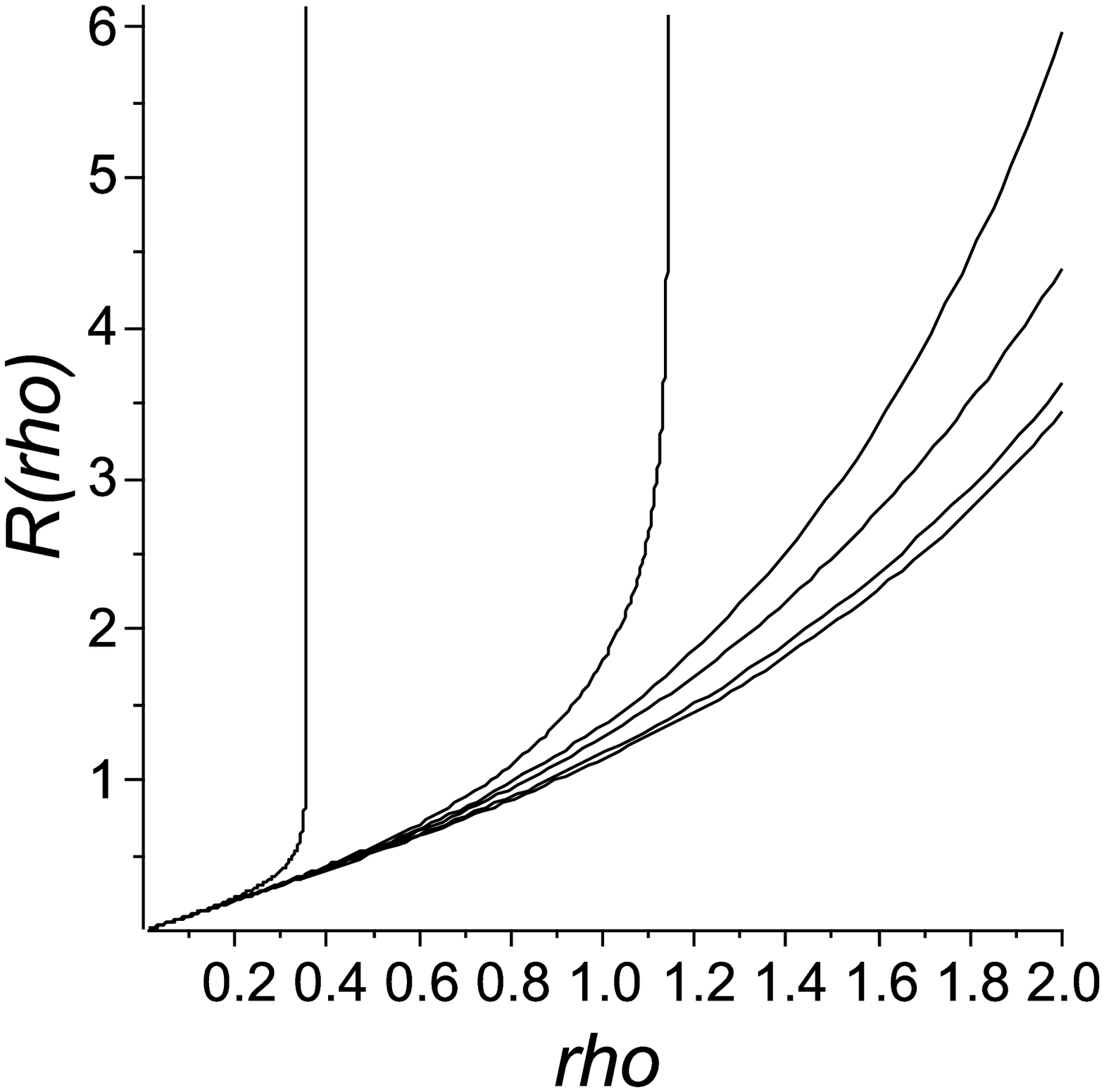}
\end{center}
\caption{Profiles of $A(\rho)$ and $R(\rho)$. 
In the left panel for $A(\rho)$, the curves from top to bottom represent the 
solutions for $k=0,-1,-3,-4,-6$ and $-32$, respectively.
For $R(\rho)$, the curves from bottom to top are for $k=0,-1,-3,-4,-6$ and $-32$.
When $k<-4$, a solution has a singularity.}
\label{fig:ar}		
\end{figure}

\begin{figure}[h]
\begin{center}
\includegraphics[width=8cm]{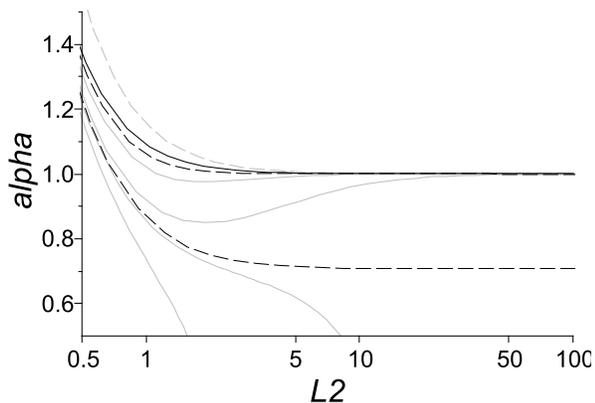}
\end{center}
\caption{$L_2 = R(\rho_0)$ dependence of $\alpha$.  $L_2$ is 
the horizon radius on the brane for each $k$. 
The dark solid line corresponds to the $k=0$ solution.
The lines that run above it, 
including the light dashed line for $k=100$,
are for $k>0$.
Those solutions for $k>0$ do not represent black holes.
The dark-dashed lines represent the solutions for $k=-1$ and  $-4$ from the top, and 
the light-solid lines represent those for $k=-2, -3.5, -4.1$ and $-5$ from the top.}	
\label{fig:aq}
\end{figure}

We show the behaviours of $A(\rho)$ and $R(\rho)$ in Fig.~\ref{fig:ar} and 
that of $\alpha$ in Fig.~\ref{fig:aq}.
Solutions for $k=0$ is not black hole solutions, while they are limiting 
solutions for black hole solution sequences with $k<0$.

From Fig.~\ref{fig:aq}, we can see that there is an upper bound on  
the four-dimensional horizon size on the brane, $L_2=R(\rho_0)$, for $\alpha>1$. 
Such an upper bound on the horizon size also appears in the ordinary general relativity
for black holes in the de Sitter universe~\cite{SNKM92,HSN94,MKNI98}.
We will examine this feature later in Sec.~\ref{Sec:dS}.

Next, we 
study the ratio between the five-dimensional and four-dimensional black hole entropies.
This ratio should become the unity if the bulk/boundary correspondence works,
and this expectation is confirmed to be correct
for the flat brane case in the large black hole limit~\cite{KR09}.
We would like to extend this study on the duality to the non-flat brane case.

\begin{figure}[t]
\begin{center}
\includegraphics[width=8cm]{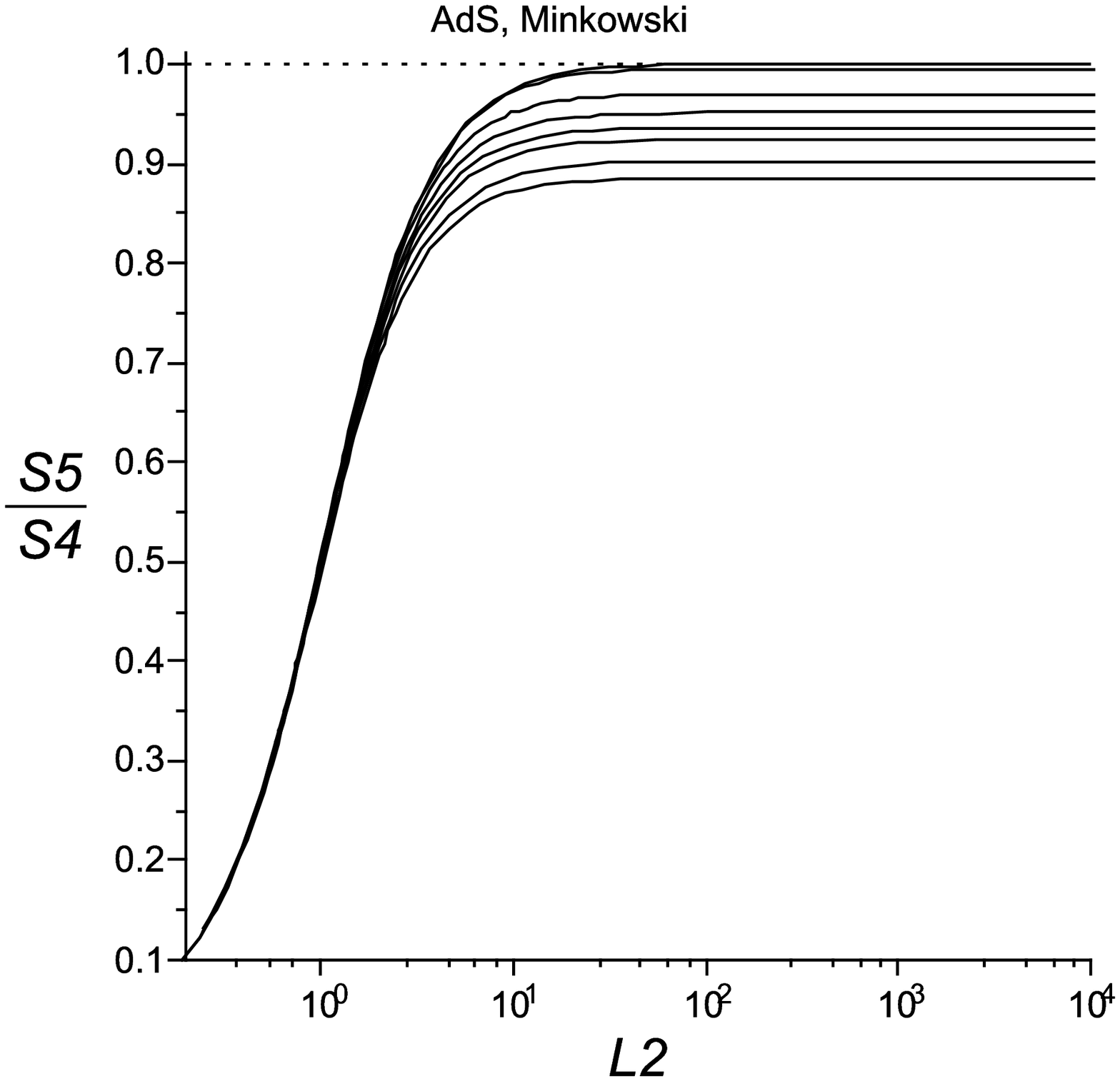}
\includegraphics[width=8cm]{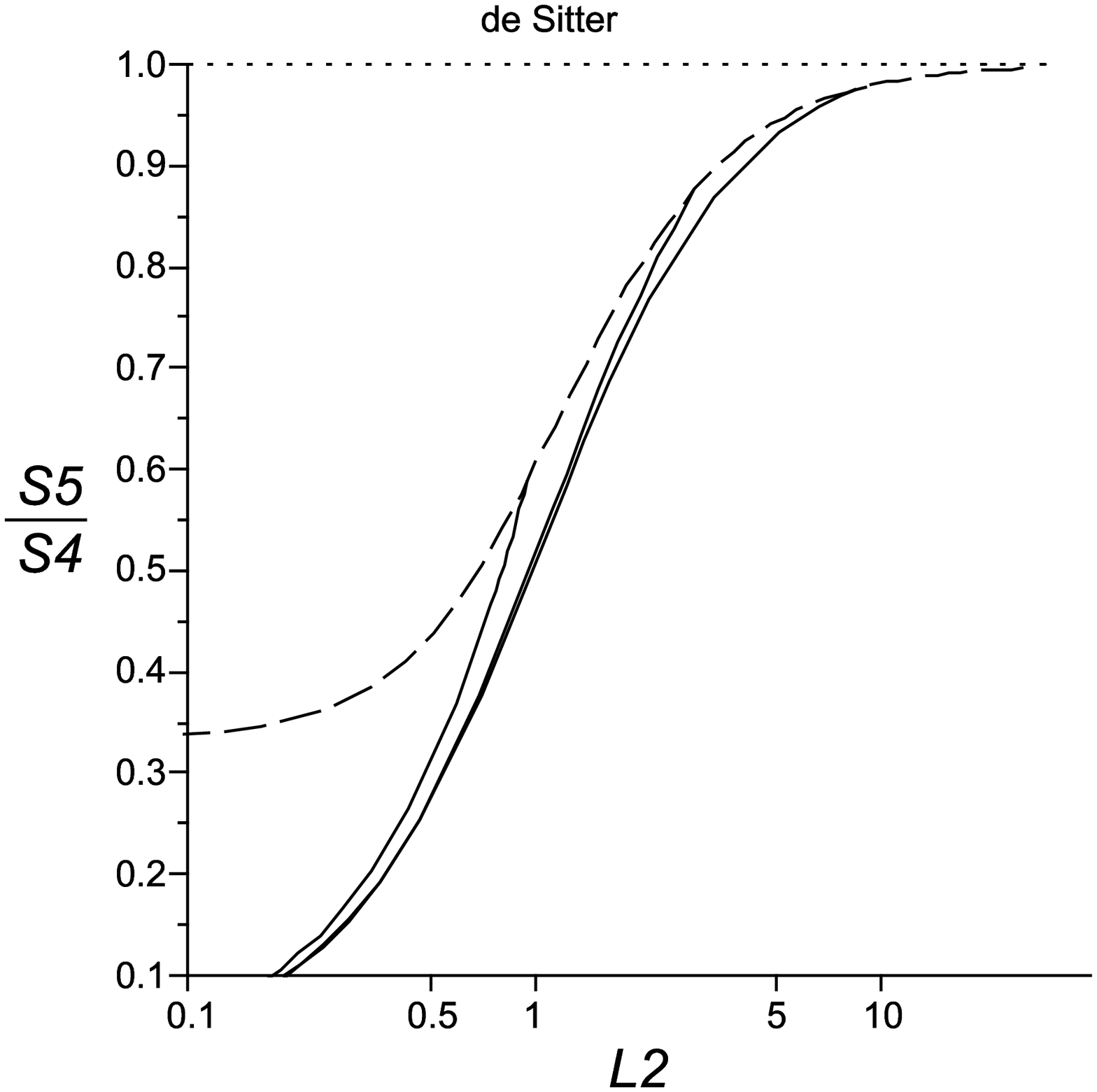}
\end{center}
\caption{$L_2=R(\rho_0)$ dependence of the entropy ratio $S_5/S_4$.
The upper panel is 
for the solutions for $\alpha \leq 1$ with asymptotically adS brane. 
Each lines correspond to 
$\alpha=1, 0.9995, 0.995, 0.99, 0.985, 0.98, 0.97$ from $0.96$ from 
top to bottom.
The lower panel is for the solutions for $\alpha > 1$ with asymptotically de Sitter or flat brane. 
The dashed line is $k=0$ case. The solid lines 
correspond to $\alpha=1.001, 1.01$ and $1.1$ from right to left.}	
\label{fig:s5s4}		
\end{figure}

The five and four-dimensional black hole entropies are defined as 
\begin{eqnarray}
S_5 = \frac{({\rm Area~of~5D~horizon})}{4G_5}
=\frac{2\pi}{G_5}\int_0^{\rho_0}R(\rho)^2d\rho
\end{eqnarray}
and
\begin{eqnarray}
S_4 = \frac{({\rm Area~of~4D~horizon})}{4G_4}=\frac{\pi}{G_4}R(\rho_0)^2,
\end{eqnarray}
respectively.
The ratio between them is given by
\begin{eqnarray}
\frac{S_5}{S_4}=\frac{G_4}{G_5}\frac{2}{R(\rho_0)^2}\int_0^{\rho_0} 
R(\rho)^2d\rho. 
\label{S5overS4}
\end{eqnarray}
We show $R(\rho_0)$ dependence of entropy ratio in Fig.~\ref{fig:s5s4}. 
The upper panel is of the solutions for $\alpha\leq 1$ with 
asymptotically adS or Minkowski branes, and the lower is 
for $\alpha > 1$ with asymptotically de Sitter branes.
As we explained in Sec.~\ref{Sec:g-coupling},
We used $G_4/G_5 =1$ in the plot for $\alpha < 1$ while $G_4/G_5=\alpha$ in the plot for $\alpha > 1$.
In the $\alpha > 1$ case, we can see that the ratio tends to the unity if 
the four-dimensional black hole radius $L_2$ is larger than the bulk curvature scale.
In the $\alpha \leq 1$ case, on the other hand, the ratio tends to some 
constant smaller than the unity as $L_2$ becomes large. 
We will study on these properties again in Secs.~\ref{Sec:dS}, \ref{Sec:AdS} and \ref{Sec:GC-for-AdS}.

\section{Large black hole limit}\label{sec:largebh}

When the black hole radius is much larger than the bulk curvature scale $l$, the brane is near 
the bulk conformal boundary and then the behaviour of the black hole on the 
brane is expected to coincide with one in the ordinary four-dimensional 
general relativity. In this section, we will discuss the large black hole 
limits with partial help of numerical analysis. 
Through out this section, we set $l=1$.

\subsection{Some basics: metric and extrinsic curvature}

\begin{figure}[t]
\begin{center}
\includegraphics[width=7cm]{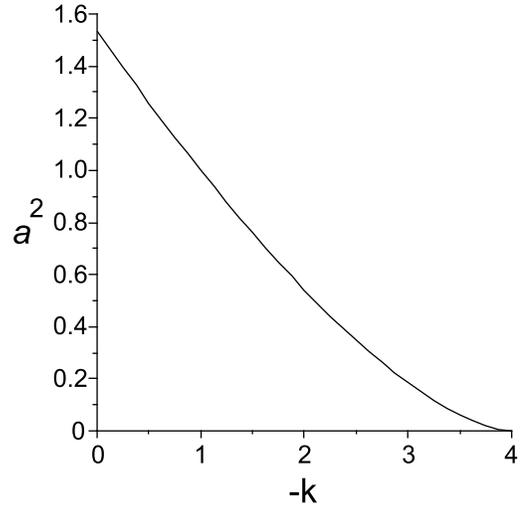}
\end{center}
\caption{$k$ dependence of $a^2$.
$a^2$ converges to a finite value $a(k=0)\sim 1.53419$ as $k\to 0$, 
becomes unity for $k=-1$ as suggested by Eq.~(\ref{exactsol1}),
and approaches zero for $k\to -4$. 
}
\label{fig:rratio}
\end{figure}

When $k > -4$, both $A(\rho)$ and $R(\rho)$ behave like $e^{\rho}$ for large $\rho$,
as we showed in Sec.~\ref{sec:solutions}.
So we can write 
$A \rightarrow A_\infty e^{\rho}$, $R \rightarrow R_\infty e^{\rho}$.
$A_\infty$ and $R_\infty$ are determined by solving the equations for each $k$. 
Then, the metric becomes
\begin{eqnarray}
ds^2 \simeq d\rho^2 + e^{2\rho}R_\infty^2(a^2d\Sigma^2+d\Omega^2), 
\end{eqnarray}
where $a \equiv A_\infty/R_\infty$, which is a function of $k$.
Note that we can know its function form
only after solving the bulk equation numerically from $\rho=0$ to $\rho=\rho_0$.
Fig.~\ref{fig:rratio} shows the $k$ dependence of 
$a^2$.
$a^2$ converges to a positive constant for $k\to 0$ and approaches zero as 
$k\to -4$.

Let us introduce a new convenient coordinate defined by $r\equiv r_0 e^{-2\rho}$ to solve 
the five-dimensional Einstein equation approximately. In this 
coordinate, the conformal boundary is at $r=0$. The metric is written as 
\begin{eqnarray}
ds^2 &=& \frac{dr^2}{4r^2} +A(r)^2d\Sigma+R(r)^2d\Omega^2 \nonumber \\
&\simeq& \frac{dr^2}{4r^2}+\frac{R_\infty^2 r_0}{r}(a^2d\Sigma^2 + d\Omega^2). 
\end{eqnarray}
Hereafter, we take $r_0=R_\infty^{-2}$ for convenience and 
we will focus on $r =\eps\ll 1$ limit. Following Ref.~\cite{HSS01}, 
we obtain the analytic solutions for $A(r)$ and $R(r)$ near 
the conformal boundary as 
\begin{eqnarray}
A(r)^2&=& \frac{a^2}{r} \Biggl[ 1+ \Bigl(\fr{6}-\frac{k}{3a^2}\Bigr)r 
-\fr{48}\Bigl(1-\frac{k^2}{a^4}\Bigr)
r^2 \log r 
\nonumber \\
&&+\Bigl(\frac{5}{288}-\frac{k}{36a^2}+\frac{5k^2}{288a^4}+\lambda\Bigr)r^2 
+ \cdots \Biggr] 
\label{ar}
\end{eqnarray}
and
\begin{eqnarray}
R(r)^2&=&\frac{1}{r} \Biggl[ 1-\left(\fr{3}-\frac{k}{6a^2}\right)r 
+\fr{48}\left(1-\frac{k^2}{a^4}\right)
r^2 \log r
 \nonumber \\
&&+\left(\frac{5}{288}-\frac{k}{36a^2}+\frac{5k^2}{288a^4}-\lambda \right)r^2 
+\dots \Biggr], \label{eq:ARexpand}
\end{eqnarray}
where $\lambda$ is an integral constant determined by $k$.
Then, the extrinsic curvature
\begin{eqnarray}
K_{\mu\nu}dx^\mu dx^\nu = K_1d\Sigma^2+K_2d\Omega^2 
\end{eqnarray}
is computed as 
\begin{eqnarray}
K_1&=&a^2\Biggl[ \fr{r} + \fr{48}\left(1-\frac{k^2}{a^4}\right)r\log r 
+ \Biggl(\frac{1}{48} \left(1-\frac{k^2}{a^4}\right)   \nonumber\\
&&- \frac{5}{288}+\frac{k}{36a^2}-\frac{5k^2}{288a^4}-\lambda \Biggr)r
+\cdots\Biggr]\label{k1exp}
\end{eqnarray}
and
\begin{eqnarray}
K_2&=&\frac{1}{r} - \fr{48}\left(1-\frac{k^2}{a^4}\right)r \log r + 
\Biggl(-\fr{48} \left(1-\frac{k^2}{a^4}\right)  \nonumber \\
&&- \frac{5}{288}+\frac{k}{36a^2}-\frac{5k^2}{288a^4}+\lambda\Biggr) r 
+\cdots. \label{eq:k1k2expand}
\end{eqnarray}
Using $K_1$ and $K_2$, 
Eqs.~(\ref{eq:jc-alpha}) and (\ref{eq:jc-Q}) are rewritten as 
\begin{eqnarray}
\alpha=\fr{2}\left(\frac{K_1}{A(\epsilon)^2}+\frac{K_2}{R(\epsilon)^2}\right) \label{eq:2.3.40}
\end{eqnarray}
and
\begin{eqnarray}
Q^2=\frac{\kappa^2_4}{\kappa^2_5} R(\epsilon)^4\left(\frac{K_2}{R(\epsilon)^2}-\frac{K_1}{A(\epsilon)^2}\right). \label{eq:2.3.50}
\end{eqnarray}

\subsection{$\alpha >1$ case:~de Sitter brane}
\label{Sec:dS}

In $\alpha>1$ case, 
positive cosmological constant is 
induced on the brane
and
the brane geometry becomes asymptotically 
de Sitter spacetime. 
From Fig.~\ref{fig:aq}, we can see that there is a restriction on the black hole 
size in the sense that the black hole size $R(\rho_0)$ has an upper bound which 
depends on $\alpha$.
The size of the black hole horizon in de Sitter spacetime 
is known to be restricted by the cosmological constant in the ordinary general 
relativity~\cite{SNKM92,HSN94,MKNI98}. From our result, we can confirm that the 
same restriction holds even in the braneworld setup. 

\begin{figure}[h]
\begin{center}
\includegraphics[width=6cm]{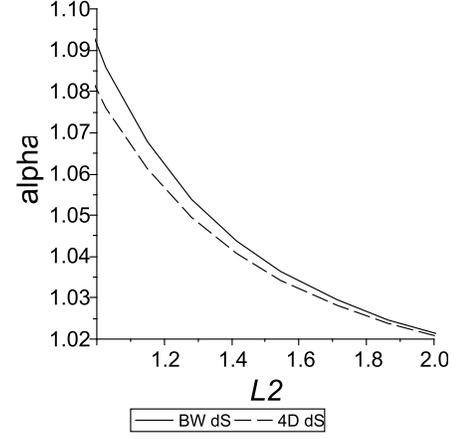}
\end{center}
\caption{Dependence of $\alpha^{\rm BW}_{\rm max}$ (solid line) 
and $\alpha^{\rm 4D}_{\rm max}$ (dashed line) about the horizon radius $L_2$.}	
\label{fig:dsbw-4d}		
\end{figure}

Comparing the braneworld upper limit $\alpha^{\rm BW}_{\rm max}$ with 
the upper limit $\alpha^{\rm 4D}_{\rm max}$ in the ordinary four-dimensional general relativity 
(see Fig.~\ref{fig:dsbw-4d}), we can see that 
\begin{eqnarray}
\alpha^{\rm BW}_{\rm max} > \alpha^{\rm 4D}_{\rm max}=\sqrt{1+\fr{6L_2^2}} \label{twoalpha}
\end{eqnarray}
is satisfied. 
It tells us that the restriction on the black hole size is weaker in the 
braneworld model.
The value of $\alpha^{\rm BW}_{\rm max}$ is given by $k=0$ solution. 

Let us study the difference between $\alpha^{\rm BW}_{\rm max}$ and 
$\alpha^{\rm 4D}_{\rm max}$ in more detail. First of all, we check 
that $\alpha_{\rm max}^{\rm 4D}$ is smaller than $\alpha^{\rm BW}_{\rm max}$. 
To focus on $\alpha^{\rm BW}_{\rm max}$, we set $k$ to zero.
Then, we see 
from Eqs.~(\ref{ar}), (\ref{eq:ARexpand}),
(\ref{k1exp}) and (\ref{eq:k1k2expand})
that 
\begin{equation}
\frac{K_1}{A^2}=1-\fr{6}\epsilon+\fr{24}\epsilon^2 \log \epsilon 
+\left(\fr{72}-2\lambda\right) \epsilon^2+
\mathcal{O}(\epsilon^3\log\epsilon)
\label{K1overA2}
\end{equation}
and
\begin{equation}
\frac{K_2}{R^2}=1+\fr{3}\epsilon-\fr{24}\epsilon^2 \log \epsilon 
+ \left(\fr{18}+2\lambda\right) \epsilon^2+
\mathcal{O}(\epsilon^3\log\epsilon). 
\label{K2overR2}
\end{equation}
Then, Eqs.~(\ref{eq:2.3.40}) yields
\begin{equation}
\alpha_{\rm max}^{\rm BW}=1+\fr{12}\epsilon+\frac{5}{144}\epsilon^2+\ord{\epsilon^3\log\epsilon}.
 \label{eq:ds-alpha} 
\end{equation}
To find the expression for $\alpha_{\rm max}^{\rm 4D}$, we should find that for 
$L_2$ first.
Since $-k/a^2\to 0$ for $k\to 0$, we find from Eq.~(\ref{eq:ARexpand}) that
\begin{eqnarray}
L_2^2=R^2(\rho_0)=\fr{\epsilon} - \fr{3} + \fr{48}\epsilon \log \epsilon
+\ord{\epsilon}. \label{l22}
\end{eqnarray}
Replacing  $L_2$ by $\epsilon$ in Eq. (\ref{twoalpha}), 
we find the expression of $\alpha_{\rm max}^{\rm 4D}$ as 
\begin{eqnarray}
\alpha_{\rm max}^{\rm 4D}=
\!
{\sqrt {1+\frac{1}{6L_2^2}}}
\!
=1+\fr{12}\epsilon 
+ \frac{7}{288}\epsilon^2 +
\mathcal{O}\left(\epsilon^3\log\epsilon\right).~~ 
\label{eq:ds-alpha-exp}
\end{eqnarray}
Then, we can confirm that $\alpha_{\rm max}^{\rm 4D}$ is smaller than $\alpha^{\rm BW}_{\rm max}$:
\begin{equation}
\alpha_{\rm max}^{\rm BW}
-\alpha_{\rm max}^{\rm 4D}
=\frac{1}{96}\epsilon^2+
\mathcal{O}(\epsilon^3\log\epsilon).
\end{equation}

Next, we  give an interpretation of the difference between them using 
the effective Einstein equations~\cite{SMS00}. 
The trace of effective Einstein equations becomes
\begin{eqnarray}
-{}^{(4)}R=-4\Lambda_4 +\frac{Q^4}{(\alpha_{\rm max}^{\rm BW})^2L_2^8} 
\label{trace}
\end{eqnarray}
and we can see that the non-linear term,
$Q^4/(\alpha_{\rm max}^{\rm BW})^2L_2^8$,
weakens the effect of the 
cosmological constant. 
This non-linear term is evaluated 
in terms of $\epsilon$
as follows.
From Eqs.~(\ref{K1overA2}) and (\ref{K2overR2}), we find that $Q^2$
of Eq.~(\ref{eq:2.3.50}) becomes
\begin{eqnarray}
Q^2=\fr{2\epsilon}-\fr{12}\log \epsilon 
+\ord{1}.
\label{Q2_ser}
\end{eqnarray}
Then, we find from Eqs.~(\ref{eq:ds-alpha}), (\ref{l22}) and (\ref{Q2_ser}) that 
\begin{eqnarray}
\frac{Q^2}{\alpha_{\rm max}^{\rm BW}L_2^4}=\fr{2}\epsilon-\fr{12}\epsilon^2 \log \epsilon
+\ord{\epsilon^2}. \label{eq:ds-Q}
\end{eqnarray}
From Eq.~(\ref{trace}), we can read off the difference between 
$\Lambda_4^{\rm 4D}$ and $\Lambda_4^{\rm BW}$ as  
\begin{eqnarray}
\delta \Lambda_4 
\simeq  - \frac{Q^4}{4(\alpha_{\rm max}^{\rm BW})^2L_2^8} \simeq - \fr{16L_2^4}.
\end{eqnarray}
In the above, we used Eqs.~(\ref{l22}) and (\ref{eq:ds-Q}). 
This matches with the value that is evaluated from 
from Eqs.~(\ref{eq:ds-alpha}) and (\ref{eq:ds-alpha-exp}) at the leading order, which is given as 
\begin{eqnarray}
\Lambda_4^{\rm 4D}-\Lambda_4^{\rm BW}=3\Bigl( (\alpha_{\rm max}^{\rm 4D})^2 
- (\alpha_{\rm max}^{\rm BW})^2 \Bigr) \simeq -\fr{16L_2^4}. 
\end{eqnarray}

\subsection{$\alpha<1$ case:~anti-de Sitter brane}
\label{Sec:AdS}

In this subsection, we consider $\alpha<1$ case in which the brane geometry 
is asymptotically anti-de Sitter spacetime. This is the so called Karch-Randall (KR) 
model~\cite{KR01}. 
In this model, 
it is expected that the relation $G_4/G_5=1$ holds approximately
when the four-dimensional adS curvature radius 
$L$ is sufficiently larger than the bulk curvature scale. 
In this section, we fix $G_4/G_5=1$ for any $L$ though we guess that this 
relation does not hold in general.
We will address this issue later in Sec.~\ref{Sec:GC-for-AdS}.

\begin{figure}[h]
\begin{center}
\includegraphics[width=8cm,height=6cm]{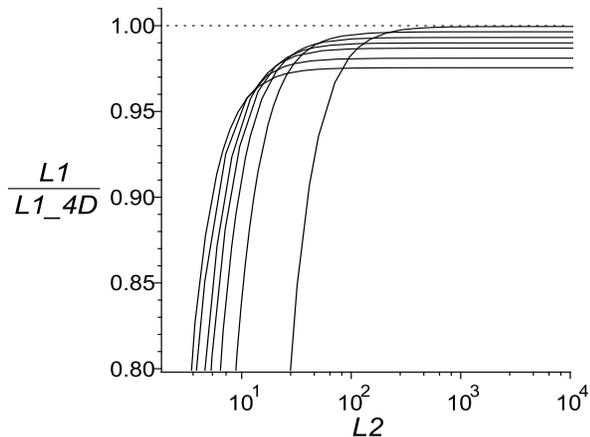}
\end{center}
\caption{$L_2$ dependence of $L_1 = |k|^{-1/2}A(\rho_0)$ for fixed values of $\alpha$. 
$L_1$ is normalised by $L_{1(\text{4D})}$.
The lines from top to bottom at large $L_2$ regime are 
for $\alpha = 0.9995,0.995,0.99,0.985,0.98,0.97$ and $0.96$, respectively.}	
\label{fig:A-alpha}		
\end{figure}

We compare the braneworld solution with four-dimensional 
extreme anti-de Sitter Reissner-Nordstr\"om (adS-RN) solution in the general relativity 
which share the same horizon radius $L_2$ and four-dimensional cosmological constant $\Lambda_4$. 
For sufficiently large 
$L_2$, 
the ${\rm adS}_2$ radius for this adS-RN solution, $L_{1(\text{4D})}$, becomes 
(see Appendix~\ref{App:RN})
\begin{eqnarray}
L_{1(\text{4D})}{}^2
=
\frac{L_2^2}{1-2\Lambda_4 L_2^2} \simeq  \fr{-2\Lambda_4}=\fr{6(1-\alpha^2)}.
\label{L4D}
\end{eqnarray}
Fig.~\ref{fig:A-alpha} shows $L_2$ dependence of $L_1 = |k|^{-1/2}A(\rho_0)$
for the braneworld black hole solutions.
From this, we can see that 
the size of ${\rm adS}_2$, $L_1$, tends to the values of four-dimensional 
adS-RN black hole, $L_{1(\text{4D})}$,
when $\Lambda_4$ is sufficiently close to zero.

\begin{figure}[h]
\begin{center}
\includegraphics[width=8cm,height=6cm]{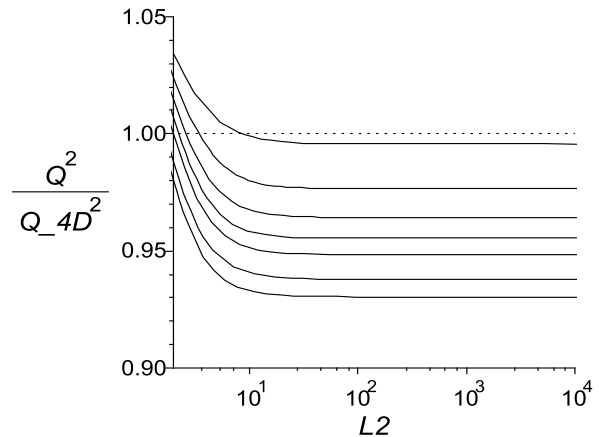}
\end{center}
\caption{$L_2$ dependence of 
$Q^2/Q^2_{4D}$. 
The lines from top to bottom are 
for $\alpha = 0.9995,0.995,0.99,0.985,0.98,0.97$ and $0.96$, respectively.}
\label{fig:Q-Qads}		
\end{figure}

Charge $Q$ for the extreme adS-RN black hole 
in the general relativity is given 
in terms of $L_2$ 
as
\begin{eqnarray}
Q_{\rm 4D}{}^2
=-L_2^4\Lambda_4
+ L_2^2.
\label{eq:adsrnq}
\end{eqnarray}
Let us compare it with that of the braneworld black hole. 
We show 
$L_2$ dependence of $Q/Q_{\text{4D}}$ in 
Fig.~\ref{fig:Q-Qads}. 
In this figure,
we see that $Q$ and $Q_{\text{4D}}$
coincide when $\Lambda_4$ is sufficiently close to zero. 
Thus, the large braneworld extreme black hole has the same near-horizon 
geometry as four-dimensional adS-RN black hole 
in the limit of vanishing $\Lambda_4$.
This figure also suggests that the 
discrepancy between $Q$ and $Q_{\text{4D}}$
can be non-zero when $\Lambda_4$ is non-zero,
and it 
becomes a constant independent of $L_2$.

Using the large black hole limit,
we can evaluate  
$L_1$ and $Q$
in terms of $1-\alpha$, that is, in terms of $\Lambda_4$.
We consider the case that both of $ l\ll L$ and $ l\ll L_2$ holds, 
where $ l$ and $L\equiv(-3/\Lambda_4)^{1/2}$ are
five and four-dimensional curvature scales
and $L_2$ is four-dimensional horizon size on the brane.
We expect that the bulk/brane duality would work under these conditions. 
In the following, we focus on $L\ll L_2$ regime
(see Appendix~\ref{App:small} for the results in $L\gg L_2$ regime).

In the $L\ll L_2$ regime, 
$L_1\sim\mathcal{O}\left((1-\alpha)^{-1}\right)$ holds
as we can see in 
Fig.~\ref{fig:A-alpha} and Eq.~(\ref{L4D}).
Then, 
from Eqs.~(\ref{ar}),
(\ref{eq:ARexpand}) and the definitions of $L_1$ and $L_2$,
we find $\epsilon\ll -k\epsilon/a^2\sim1-\alpha\ll 1$.
This regime is realized in the limit of $k\to -4$, for which $a$ tends to zero 
as shown in Fig.~\ref{fig:rratio}.
After some 
calculations in this regime, we find
for $\eps\to 0$ that~(see Appendix~\ref{App:large} for derivations of the 
following equations)
\begin{align}
\frac{L_1^2}{L_{1\text{(4D)}}{}^2} = 
1 - \frac{3}{2} (1-\alpha)+ \mathcal{O}\bigl((1-\alpha)^2\log(1-\alpha)\bigr).
\label{l1correction}
\end{align}
From this equation, as $\alpha \to 1$, we can see that $L_1^2$ approaches that of the 
four-dimensional extreme adS-RN solution, which is given by Eq.~(\ref{L4D}).

We can analyze behaviour of the charge $Q$ in the same way. 
The result is 
\begin{equation}
\frac{Q^2}{Q_{\rm 4D}{}^2}
=1+2(1-\alpha)\log(1-\alpha)+\mathcal{O}(1-\alpha),
\label{eq:qcorrection}
\end{equation}
and we find that $Q$ approaches that of
the four-dimensional adS-RN solution, Eq.~(\ref{eq:adsrnq}), 
in the limit of $\alpha \to 1$.

\subsection{Gravitational coupling for adS branes}
\label{Sec:GC-for-AdS}

If the adS/CFT correspondence holds in the KR model, it is natural to expect that 
$S_5=S_4$ holds, at least in the large black hole limit. 
In this paper, however, we observed that $S_5 \neq S_4$ in that limit
when we suppose $G_4/G_5=1$.
In this subsection, we would like to propose a formula for $G_4/G_5$ 
which makes $S_5$ equal to $S_4$ for any $\Lambda_4$.

\begin{figure}[h]
\begin{center}
\includegraphics[width=7cm]{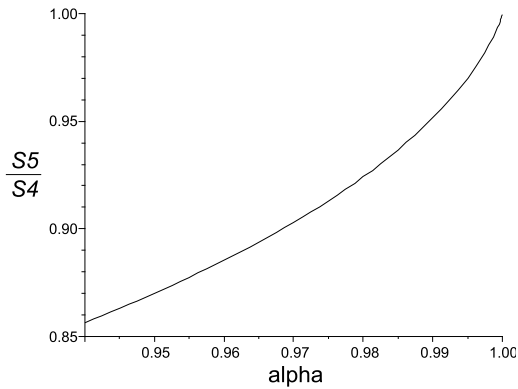}
\end{center}
\caption{$\alpha$ dependence of the entropy ratio $S_5/S_4$ calculated in the 
large black hole limit $\rho_0\to\infty$.}
\label{fig:s5s4lim}		
\end{figure}

In Fig.~\ref{fig:s5s4lim},
we show $\alpha$ dependence of the ratio $S_5/S_4$ in the large black hole limit.
This entropy ratio is proportional to $G_4/G_5$ as shown in Eq.~(\ref{S5overS4}).
Then, the value of $G_4/G_5$ that makes the entropy ratio to be unity will be 
inverse of the value of $S_5/S_4$ shown in Fig.~\ref{fig:s5s4lim}.

Let us study this value of $G_4/G_5$ in the limit of 
$\alpha \to 1$.
In this limit,
we can expand $A_5$ as~(see Appendix~\ref{App:large})
\begin{align}
A_5=\frac{4\pi}{\eps}\Bigl(
1 + 2(1-\alpha)\log(1-\alpha) + \mathcal{O}\left(1-\alpha\right)
\Bigr).
\label{eq:5darea} 
\end{align}
Then, we obtain
\begin{align}
\frac{S_5}{S_4}&=
\frac{G_4}{G_5}
\frac{A_5}{A_4}
\notag \\
&=
\frac{G_4}{G_5}
\Bigl(
1+2\left(1-\alpha\right)\log\left(1-\alpha\right)+\ord{1-\alpha}
\Bigr). 
\label{eq:entropyratio}
\end{align}
In order that this ratio is equal to unity, we should set $G_5/G_4$ as 
\begin{align}
\frac{G_5}{G_4} &= 1 + 2(1-\alpha)\log(1-\alpha)+\cdots
\notag \\
&= 1 + \frac{l^2}{L^2}\log\left(\frac{2l^2}{L^2}\right) + \cdots.
\label{Gratio}
\end{align}

Since the charge $Q$ is also proportional to $G_4/G_5$,
as seen in Eq.~(\ref{eq:jc-Q}),
we may determine the value of $G_4/G_5$ requiring that the charge ratio 
$Q/Q_{4D}$ becomes unity as $\alpha \to 1$. 
Interestingly, the expression of $G_4/G_5$ determined in this way coincides
with Eq.~(\ref{Gratio}) at least up to sub-leading order in the $\alpha\to 1$ limit (see Eq.~(\ref{eq:qcorrection})). 
This fact may imply that $G_4$ and $G_5$ should be related by Eq.~(\ref{Gratio}) for 
general $\alpha$, that is, for general $\Lambda_4$.


\section{Summery and Discussion}\label{sec:summery}

In this paper,
we analyzed the near-horizon geometry of charged extreme 
black holes localized on the brane with non-vanishing cosmological constant
in the RS-type braneworld models.
In the de Sitter brane case, 
we find that there is an upper bound on the black hole size
and that the bound is determined by the cosmological constant on the brane.
This restriction on the horizon size also appeared
 in the ordinary four-dimensional general relativity, while 
the restriction was found to be weaker in the braneworld case
due to the non-linear term in the effective Einstein equation on the brane.

In the anti-de Sitter brane case, 
we observed discrepancies between
the near-horizon geometry of the brane-localized black hole
from that of the four-dimensional extreme adS-RN black 
hole. 
We found that 
the ${\rm adS}_2$ radius 
and the charge are 
smaller than those of four-dimensional adS-RN black holes, and confirmed that 
those discrepancies vanish in the flat brane limit ($\alpha\to 1$).
We also calculated the five and four-dimensional black hole entropies 
assuming $G_4/G_5=1$. As a result, it turned out that 
$S_5/S_4$ becomes smaller as the cosmological constant on the brane becomes larger. 

In the Karch-Randall model, 
it is suggested by Ref.~\cite{KS05} that 
the four-dimensional gravity weakens as
$G_4/G_5 \approx 1 - \ord{l^2/L^3}{\mathcal R}$ for
$L\lesssim\mathcal{R}\lesssim L^3/l^2$,
where $\mathcal R$ is the separation of two gravitating objects,
due to small four-dimensional graviton mass.
However, the formula for $G_4/G_5$ we proposed in this paper, Eq.~(\ref{Gratio}), 
has a different form from it. 
It may be peculiar that
our formula is 
independent of the black hole size, while the formula of Ref.~\cite{KS05} depends 
on propagation distance $\mathcal R$ of the gravity.
It will be interesting to investigate whether these two 
formulae are compatible or not.

There are many remaining issues. In our work, we addressed the near-horizon geometry only. 
To justify our result, 
we have to construct the full bulk solutions.
Perturbative approaches for the solution construction
like Ref.~\cite{Kod08} 
or numerical solution construction methods like Ref.~\cite{KTN03}
may give fruitful results.
In Ref.~\cite{Yos09}, it is pointed out that nonsystematic error increases as
 taking the asymptotic boundary farther from the horizon
 even if the black hole radius is smaller than the AdS curvature radius, 
 which could imply the singularity formation in the bulk.
It is also valuable to examine whether such nonsystematic error exist in the extremal case.
Another interesting subject is the near-horizon geometry of 
a rotating extreme black hole localized on the brane. Such a black hole 
has the spontaneous emission through the superradiant modes \cite{sm}, although its 
temperature is zero. Thus, the adS/CFT correspondence about the braneworld models may suggest  
that such a black hole will be a dynamical. It is meaningful to address if it is true 
or not. These studies will be helpful for understanding on the black hole solutions in 
higher-dimensional spacetime models and also on the adS/CFT 
correspondence in generalized situations.

\begin{acknowledgements}
We would like to thank Roberto Emparan, Shunichiro Kinoshita, Takashi Nakamura and Takahiro Tanaka 
for useful discussions. This work was supported by the Grant-in-Aid for the 
Global COE Program"The Next Generation of Physics, Spun from Universality and Emergence"
from the Ministry of Education, Culture, Sports, Science and Technology
(MEXT) of Japan. TS is partially supported by Grant-Aid for Scientific Research from Ministry of 
Education, Science, Sports and Culture of Japan (Nos.~21244033,~21111006,~20540258 
and 19GS0219), the Japan-U.K. Research Cooperative Programs. 
\end{acknowledgements}

\appendix
\section{4D Reissner-Nordstr\"om black hole}
\label{App:RN}

In this appendix, we summarise the fundamental features of the extreme static charged black hole 
solutions in the four-dimensional ordinary general relativity.
We use this solution as a fiducial to compare with the 
brane-localized charged black holes in this paper.

The metric of charged black hole solutions in the four-dimensional ordinary general relativity 
with a cosmological constant $\Lambda_4$ is given as 
\begin{eqnarray}
ds^2=-f(r)dt^2+\frac{dr^2}{f(r)}+r^2d\Omega^2,
\end{eqnarray}
where
\begin{eqnarray}
f(r)=1-\frac{\Lambda_4}{3}r^2+\frac{Q^2}{r^2}-\frac{2M}{r}. 
\end{eqnarray}
The horizon radius $r_H$ is determined by $f(r_H)=0$,
which implies
\begin{eqnarray}
M=\fr{2}\Bigl(r_H+\frac{Q^2}{r_H}-\frac{\Lambda_4}{3}r_H^3\Bigr).
\end{eqnarray}
When the black hole is extreme, $f'(r_H)=0$ holds.
In this case, we find
\begin{eqnarray}
\Lambda_4r_H^4-r_H^2+Q^2=0.
\end{eqnarray}
One of roots for this is given by 
\begin{equation}
r_H^2=\fr{2\Lambda_4}\left(1-\sqrt{1-4\Lambda_4Q^2}\right),
\end{equation}
and $r=r_H$ determined by this equation will be the black hole horizon.
Now, $f(r)$ is written as 
\begin{equation}
f(r)=(r-r_H)^2 \times \frac{g(r)}{r^2},
\end{equation}
where 
\begin{equation}
g(r) \equiv 1-\frac{\Lambda_4}{3}\left(r^2+2r_Hr+3r_H^2\right).
\end{equation}

If $\Lambda_4>0$, the equation $f'(r)=0$ 
has another positive root $r=\tilde r_H$, 
which is given by 
\begin{equation}
\tilde r_H^2 = \fr{2\Lambda_4}\left(1+\sqrt{1-4\Lambda_4Q^2}\right).
\end{equation}
The surface $r=\tilde r_H$ is, however,
not the black hole horizon because $g(\tilde r_H)<0$. 
It is rather the cosmological horizon of the de Sitter universe.

The near-horizon geometry of this extreme black hole is given by
\begin{eqnarray}
ds^2 \simeq \frac{r_H^2}{g(r_H)}\left(-x^2 d{t'}^2+\frac{dx^2}{x^2}\right)+r_H^2d\Omega^2,
\end{eqnarray}
where we 
introduced new coordinates as
$x=r-r_H$ and $t'=\frac{g(r_H)}{r_H}t$. 
As is well-known, this geometry is ${\rm adS}_2 \times {\rm S}^2$. 
The radius of each submanifold is given by 
\begin{equation}
L_1^2=\frac{r_H^2}{g(r_H)}=\frac{r_H^2}{\sqrt{1-4\Lambda_4 Q^2}}
= \frac{L_2^2}{1-2\Lambda_4 L_2^2}
\label{L14D}
\end{equation}
and
\begin{equation}
L_2^2=r_H^2. 
\end{equation}


\section{Large black hole limit in anti-de Sitter case}
\label{App:adSBH}

In this section, we give a detailed analysis on the large black hole limit in 
adS brane case of Sec.~\ref{Sec:AdS}.
We focus on a regime in which both of $L\gg l$ and $L_2\gg l$ are satisfied, i.e.,
the regime in which the adS/CFT correspondence would work,
and investigate on $L_2 \ll L$ and $L \ll L_2$ cases in 
subsections~\ref{App:small} and \ref{App:large}, respectively.
We set $ l=1$ in this section unless otherwise noted.

To facilitate the following analysis, we introduce 
\begin{equation}
\ga\equiv -\frac{k}{a^2},
\qquad
\de \equiv \gamma - 1,
\qquad 
\beta\equiv 1-\alpha.
\end{equation} 
Note that $\ga$ and $\de$ are functions of $k$.  $\de$ becomes zero for $k=-1$, increases 
monotonically as $k$ decreases, and diverges as $k\to-4$, 
as we can see from Fig.~\ref{fig:rratio}.
In the case of adS brane with $L\gg l$ and $L_2\gg l$, we may assume that $\eps$ and 
$\beta$ are positive value much smaller than the unity.
In this case, we find from Eq.~(\ref{eq:2.3.40}) that 
\begin{align}
 \beta
&=
\frac{\gamma-1}{12}\eps
 -\frac{5\ga^2+8\ga+5}{144}\eps^2
 +\mathcal{O}{\left(\ga^3\eps^3\log\eps,\eps^3\log\eps\right)}
\notag \\
&=
\frac{1}{12}\de\eps 
-\frac{5\de^2+18\de+18}{144}\eps^2
  +\mathcal{O}{\left(\de^3\eps^3\log\eps,\eps^3\log\eps\right)},
\label{eqalpha}
\end{align}
where we used Eqs.~(\ref{ar}), (\ref{eq:ARexpand}), (\ref{k1exp}) and 
(\ref{eq:k1k2expand}) and assumed $\ga\epsilon\ll 1$ so that the expansion converges.
We treat $\de\ll 1$ and $\de\gg 1$ cases separately in the following.

\subsection{$\de \ll 1$ case and $L_2 \ll L$ regime }
\label{App:small}

For $\de \ll 1$, dominant part of Eq.~(\ref{eqalpha}) is given by
\begin{equation}
 \beta=\frac{1}{12}\de\eps - \frac{1}{8}\eps^2
+ \mathcal{O}\left(\de\eps^2, \eps^3\log\eps\right).
\end{equation}
This equation in terms of $\eps$ has two roots for $\beta<\de^2/72$,
and they are given by
\begin{equation}
 \eps=\eps_\pm
\equiv
\frac{1}{3}\left(
\de \pm \sqrt{\de^2-72\be}
\right).
\end{equation}
Let us inspect $\eps_+$ first. 
It behaves as $\eps_+\sim 2\de/3$ when $\be\ll\de^2$.
Since $L_2 \simeq  l/\sqrt{\eps}$ and $L\simeq  l/\sqrt{2\be}$,
$(L_2/l)^2 \ll L/l$ and thus $L_2 \ll L$ follows in this regime, that is, the four-dimensional black hole 
radius on the brane becomes much smaller than the four-dimensional curvature 
scale when $\be\ll\de^2$.
In this regime, it is convenient to parametrize the deviation of $\eps_+$ 
from $2\de/3$ as 
\begin{equation}
 \de \equiv \frac{3}{2}\left(1+\chi\right)\eps,
\end{equation}
where we assume $0<\chi\ll 1$.
In this notation, $\eps$ is related to $\beta$ as
\begin{equation}
\be = \frac{1}{12}\chi\eps^2 
+ \mathcal{O}\left(\eps^3\log\eps\right).
\end{equation}
The $\chi$ term in the right-hand side will be dominant over 
$\mathcal{O}(\eps^3\log\eps)$ term if $\chi\gg\eps\log\eps$.
We find in this regime that 
the expansion forms of $L_1^2$, $L_2^2$ and $Q^2$ become
\begin{align}
L_1^2 &= 
\frac{1}{\eps} -1 -\frac{3}{2}\chi 
+ \left( \frac{33}{16} + \la +\frac{17}{4}\chi \right)\eps
+\mathcal{O}\left(\eps^2\log\eps\right),
\\
L_2^2 &= 
\frac{1}{\eps} -\frac{1}{2}
- \left( \frac{3}{16} + \la +\frac{1}{4}\chi \right)\eps
+\mathcal{O}\left(\eps^2\log\eps\right),
\\
Q^2 &=
\frac{1}{\eps} -\frac{1}{4} + 4\la + \frac{3}{4}\chi
+\mathcal{O}\left(\eps\log\eps\right).
\end{align}
$\la$ in the above is a function of $k$ and $\eps$, 
i.e., a function of $\de$ and $\eps$, and it 
is determined so that the bulk geometry becomes regular. Since we know that the 
bulk metric reduce to that of adS$_5$ in the limit of $\chi\to 0$ and $\eps\to 
0$, we can fix the leading term of $\la$ as (see also ref.~\cite{KR09})
\begin{equation}
 \la|_{\eps=0=\chi}=0.
\end{equation}
Note that $\eps$ or $\chi$ may appear in the sub-leading terms of $\la$.
Then, we find the correct expansion forms of $L_1^2$, $L_2^2$ and $Q^2$
to be
\begin{align}
 L_1^2 &= 
\frac{1}{\eps} -1 -\frac{3}{2}\chi 
+\frac{33}{16}\eps
+\mathcal{O}\left(\lambda\eps, \chi\eps, \eps^2\log\eps\right),
\\
L_2^2 &= 
\frac{1}{\eps} -\frac{1}{2} - \frac{3}{16} \eps
+\mathcal{O}\left(\lambda\eps,\chi\eps,\eps^2\log\eps\right),
\\
Q^2 &=
\frac{1}{\eps} -\frac{1}{4} 
+\mathcal{O}\left(\lambda,\chi,\eps\log\eps\right).
\end{align}
These expression coincide with those for flat brane case given in~\cite{KR09} 
in the limit of $\chi\to 0$, and 
difference appears only in $L_1$
up to the order shown here.
We have to clarify sub-leading behavior of $\la$ to fix the higher 
order terms of these expansion equations, 
while it seems difficult to do it analytically.

Next, we make some comments on another solution $\eps_-$.
Let us fix $\beta$ and consider $\de$ dependence of $\eps_-$.
Fixing $\be$, we can show that $\eps_-$ monotonically decreases as we increase $\de$, 
and $\eps_-$ takes the maximum value $\de/3$ for $\de=6\sqrt{2\be}$.
This behaviour can be expressed equivalently as $L/l \lesssim (L_2/l)^2$, that is, the 
brane black hole size is of the same order as or
larger than the four-dimensional curvature scale.
The brane black hole size $L_2$ grows as $L_2\sim 12\be/\de$ for $\de^2\gg\be$.
This branch of solution is smoothly connected to that for $\de\gg 1$, and we 
will analyze it in the next subsection in detail.

\subsection{$\de \gg 1$ case and $L \ll L_2$ regime }
\label{App:large}

We will focus on the case $\de$ and then $\ga$ is much larger than the unity in 
the aim of studying the black holes much larger than the four-dimensional 
curvature scale.
We use $\ga$ instead of $\de$ throughout this subsection.

Before solving Eq.~(\ref{eqalpha}) to find the black hole radius, we fix the 
leading behaviour of $\la$ from the bulk regularity.
From Eqs.~(\ref{eq:alpha}), (\ref{ar}), (\ref{eq:ARexpand}), (\ref{k1exp}),
(\ref{eq:k1k2expand}) and (\ref{eq:2.3.40}),
we find 
\begin{align}
\frac{L_1^2}{L_{1\text{(4D)}} {}^2}
&=
1 - \frac{1}{8}\ga\eps-\frac{1}{6}\ga^2\eps^2\log\eps
-\frac{1}{4}\eps 
-\frac{1}{8}\frac{\eps}{\ga}
\notag \\
&\qquad\qquad
- 8\la\eps^2
+ \mathcal{O}\left(
\ga^2\eps^2, \ga\eps^2\log\eps
\right),
\label{L1ratio}
\end{align}
where $L_{1\text{(4D)}}$ is the radius of four-dimensional adS-RN solution, 
which is given by Eq.~(\ref{L14D}).
For a fixed $\ga\eps$, this ratio should converge to some constant 
in the limit of $\eps\to 0$, as we can see in Fig.~\ref{fig:A-alpha},
while the third term in the right-hand side,
$\ga^2\eps^2\log\eps$, diverges in such a limit.
We have only $\la\eps^2$ term to cancel such divergence.
This $\la$ is a function of $\ga$ and $\eps$, and may have 
the following leading behaviour:
\begin{equation}
\la=\frac{\ga^2}{48}\log\ga.
\label{lambda}
\end{equation}
This $\la$ replaces the logarithmic term as $\log\eps\to\log\ga\eps$
and the divergence is canceled.
We use this leading form of $\la$ henceforth.

When $\de\gg 1$, the expression of $\be$, Eq.~(\ref{eqalpha}), can be expressed 
as 
\begin{equation}
\be=
\frac{\teps}{12}  - \frac{\eps}{12}
- \frac{5}{144}\teps^2 - \frac{1}{8}\eps\teps
+\mathcal{O}\left( \teps^3\log\eps, \eps^2, \la\teps\eps^2  \right),
\label{eqteps}
\end{equation}
where we introduced $\teps\equiv\ga\eps$. 
Note that $\eps\ll\teps\ll 1$ by assumption
and then $\beta\sim\teps\gg\eps$, i.e.,
$L_2\gg L$ follows in this regime.
Solving Eq.~(\ref{eqteps}) as an equation of $\teps$ and expanding it with 
respect to $\be$ and $\eps$, we find a solution 
which satisfies the condition $\teps\ll 1$ as 
\begin{equation}
\teps = 12\be + 60 \be^2
 +\left( 1+18\be \right) \eps 
+ \mathcal{O}\left( \be^3, \eps^2  \right).
\label{tepsform}
\end{equation}
Plugging this expression into Eq.~(\ref{L1ratio}), we find 
\begin{align}
\frac{L_1^2}{L_{1\text{(4D)}}{}^2}
&= 1 -\frac{3}{2}\be - \frac{3}{8}\eps
+ \mathcal{O}\left(\be^2\log\be, \frac{\eps^2}{\be}\right),
\end{align}
and we obtain Eq.~(\ref{l1correction}) by taking $\eps$ to zero 
while keeping $\be$ fixed.
To proceed the expansion and determine the higher-order terms, we have to know 
the sub-leading behaviour of $\la$, while it 
seems not straightforward.

In a similar manner, we can evaluate the ratio of $Q$ of a brane-localized 
black hole to $Q_{\rm 4D}$ of the four-dimensional adS-RN solution, which is 
given by Eq.~(\ref{eq:adsrnq}).
Using Eqs.~(\ref{lambda}) and (\ref{eqteps}), we obtain the expansion form of
$Q^2/Q_{\rm 4D}{}^2$, Eq.~(\ref{eq:qcorrection}), as
\begin{align}
\frac{Q^2}{Q_\text{4D}{}^2}
&=
1 + \frac{1}{6}\teps\log\teps - \frac{1}{6}\eps\log\teps 
+ \mathcal{O}\left(\teps\right)
\notag \\
&=
1 + 2 \be\log\be 
+ \mathcal{O}\left(\be,\eps\right).
\end{align}

Finally, let us calculate 
the five-dimensional horizon area 
and its ratio to the four-dimensional horizon area on the brane.
The area of five-dimensional horizon is given as
\begin{align}
A_5 &=
2\int_0^{\rho(r=\eps)} 4\pi R\left(\rho\right)^2 d\rho
\notag \\
&=
8\pi\int_0^{\ro\left(r=1/\ga\right)}
\!\!\!\!\!\!
R\left(\rho\right)^2d\rho
+
4\pi\int_\eps^{1/\ga} \frac{R(r)^2}{r} dr,
\label{A5}
\end{align}
where we divided the integral into two pieces for convenience of the following calculation.
Note that $\eps\ll1/\ga\ll 1$ by assumption.
Using $R(\rho)\simeq R_\infty^2 e^{2\rho}$ and $R^2\simeq 1/r$, 
which hold for $\rho\gg 1$ and $r\ll 1$,
the first integral 
in the right-hand side of Eq.~(\ref{A5}) becomes
\begin{equation}
8\pi\int_0^{\ro\left(r=1/\ga\right)}
\!\!\!\!\!\!
R\left(\rho\right)^2d\rho
\simeq
4\pi R^2|_{r=1/\ga}
=\mathcal{O}\left(\ga\right)
=\mathcal{O}\left(\frac{\teps}{\eps}\right).
\label{int1}
\end{equation}
Using Eqs.~(\ref{eq:ARexpand}) and (\ref{lambda}),
the second integral of Eq.~(\ref{A5}) becomes
\begin{align}
&4\pi\int_\eps^{1/\ga}\frac{R(r)^2}{r}dr
\notag \\
&=
4\pi\int_\eps^{1/\ga} dr
\left(
\frac{1}{r^2}
-\frac{2+\ga}{6r}
+ \mathcal{O}\left(\ga^2\log(\ga r),r\right)
\right)
\notag \\
&=
4\pi\left[
-\frac{1}{r}-\frac{2+\ga}{6}\log r
+\mathcal{O}\left(
\ga^2 r\log\left(\ga r\right)
\right)
\right]^{1/\ga}_{\eps}
\notag \\
&=
\frac{1}{\eps}\left(
1+\frac{2\eps+\teps}{6}\log\teps
+\mathcal{O}\left(\teps\right)
\right).
\label{int2}
\end{align}
Since the first integral, Eq.~(\ref{int1}), can be absorbed in 
$\mathcal{O}\left(\teps\right)$ of Eq.~(\ref{int2}),
we find that $A_5$ is given by Eq.~(\ref{int2}).
Writing it in terms of $\be$ and $\eps$, we obtain
\begin{equation}
A_5=\frac{4\pi}{\eps}\left(
1+2\be\log\be+\frac{1}{2}\eps\log\be
+\mathcal{O}\left(\be\right)
\right),
\label{a5final}
\end{equation}
and it gives Eq.~(\ref{eq:5darea}) for $\eps\to 0$.
It is straightforward to calculate the ratio between the five and 
four-dimensional horizon areas, $A_5/A_4=A_5/(4\pi L_2^2)$.
Using Eqs.~(\ref{a5final}) and (\ref{eq:ARexpand}), we find 
\begin{equation}
\frac{A_5}{A_4}=
1+2\be\log\be+\frac{1}{2}\eps\log\be
+\mathcal{O}\left(\be\right).
\end{equation}
This yields Eq.~(\ref{eq:entropyratio}) for $\eps\to 0$.

\end{document}